\documentclass[twocolumn,aps,prl,preprintnumbers]{revtex4}
\usepackage{bbm}
\usepackage[latin9]{inputenc}
\usepackage{amsmath}
\usepackage{amssymb}
\usepackage{graphicx}
\usepackage{mathrsfs}
\usepackage{amsfonts}
\usepackage{amsthm}
\usepackage{color}
\usepackage{txfonts}
\usepackage[colorlinks=true,citecolor=blue,linkcolor=blue,urlcolor=blue,anchorcolor=blue]{hyperref}%
\usepackage{pifont}
\hypersetup{colorlinks=true,citecolor=blue,linkcolor=blue,urlcolor=blue}
\providecommand{\U}[1]{\protect\rule{.1in}{.1in}}
\makeatletter
\@ifundefined{textcolor}{}
{
\definecolor{BLACK}{gray}{0}
\definecolor{WHITE}{gray}{1}
\definecolor{RED}{rgb}{1,0,0}
\definecolor{GREEN}{rgb}{0,1,0}
\definecolor{BLUE}{rgb}{0,0,1}
\definecolor{CYAN}{cmyk}{1,0,0,0}
\definecolor{MAGENTA}{cmyk}{0,1,0,0}
\definecolor{YELLOW}{cmyk}{0,0,1,0}
}

\makeatother
\begin{document}
\title{Nonlinear Topological Magnon Spin Hall Effect}
\author{Zhejunyu Jin$^1$}
\author{Xianglong Yao$^1$}
\author{Zhenyu Wang$^1$}
\author{H. Y. Yuan$^2$}
\author{Zhaozhuo Zeng$^1$}
\author{Yunshan Cao$^1$}
\author{Peng Yan$^1$}
\email[]{yan@uestc.edu.cn}
\affiliation{$^1$School of Electronic Science and Engineering and State Key Laboratory of Electronic Thin Films and Integrated Devices, University of
Electronic Science and Technology of China, Chengdu 610054, China\\$^2$Institute for Theoretical Physics, Utrecht University, 3584 CC Utrecht, The Netherlands}

\begin{abstract}

When a magnon passes through two-dimensional magnetic textures, it will experience a fictitious magnetic field originating from the $3\times 3$ skew-symmetric gauge fields. To date, only one of the three independent components of the gauge fields has been found to play a role in generating the fictitious magnetic field while the rest two are perfectly hidden. In this work, we show that they are concealed in the nonlinear magnon transport in magnetic textures. Without loss of generality, we theoretically study the nonlinear magnon-skyrmion interaction in antiferromagnets. By analyzing the scattering features of three-magnon processes between the circularly-polarized incident magnon and breathing skyrmion, we predict a giant Hall angle of both the confluence and splitting modes. Furthermore, we find that the Hall angle reverses its sign when one switches the handedness of the incident magnons. We dub it nonlinear topological magnon spin Hall effect. Our findings are deeply rooted in the bosonic nature of magnons that the particle number is not conserved, which has no counterpart in low-energy fermionic systems, and may open the door for probing gauge fields by nonlinear means.
\end{abstract}

\maketitle

\textit{Introduction.---}Topology dictates the particle or wave transport in many branches of physics, ranging from solid state physics to geophysics and astrophysics \cite{Delplace2017,Parker2020}. One outstanding example in condensed matter physics is the intrinsic spin Hall effect which originates from the momentum-space topology, i.e., the Berry curvature of the band structure \cite{Kato2004,Wunderlich2005,Sinova2015,Kavokin2005,Leyder2007,Onose2010,Sheng2020,Sinova2004,Guo2008}. On the other hand, non-collinear spin textures, such as the magnetic vortex, meron, and skyrmion, can give rise to the real-space topology. When a spinful particle propagates through the topological spin texture, it will experience an effective Lorentz force, resulting in the so-called topological (spin-) Hall effect \cite{Yin2015,Mook2017,Akosa2018,Daniels2019,Kim2019,Du2022}.

Magnons, quanta of spin waves, are the collective excitations of ordered magnets \cite{Yan2011,Yuan2022}. Very recently, magnon-based spintronics has attracted enormous interest due to peculiar advantages of magnons, such as the long-distance transport and low-energy consumption. Magnons carry spin angular momentum as well, so that they can experience an effective Lorentz force from the spin texture, leading to the topological magnon Hall effect \cite{Dugaev2005,Kovalev2012,Hoogdalem2013,rainbow,Lan2021}. In antiferromagnets, magnons have two degenerate modes with opposite spins, i.e., right- and left-handed magnons \cite{Wangst2022}. Therefore, when a magnon passes through the antiferromagnetic (AFM) skyrmion, for instance, it will experience a spin-dependent Lorentz force, resulting in the topological magnon spin Hall effect \cite{Daniels2019,Jin2021,Liu2022}. These topological magnon Hall effects originate from the gauge fields in transforming the non-collinear magnetic texture to the collinear state. The gauge transformation generates the covariant form of the differential operator $ \partial_{\mu} + \mathcal{A}_{\mu}$ with $\mu=x,y$. Here, the $3\times3$ skew-symmetric matrix $\mathcal{A}_{\mu}=\mathcal{R}^{-1}\partial_\mu \mathcal{R}$ with the rotation matrix $\mathcal{R}$ contains three independent gauge fields \cite{Tan2020,Tatara2019}. So far, only one of the three elements, i.e., $\mathcal{A}_{\mu,12}$, has been identified to play a role in the Hall transport of magnons while the rest two ($\mathcal{A}_{\mu,13}$ and $\mathcal{A}_{\mu,23}$) are concealed from the community.

In the past few years, the nonlinear Hall effect due to the momentum-space topology, e.g., Berry curvature dipole, has attracted much attention \cite{Sodemann2015,Ma2019,Kang2019,Du2018,He2019,Shao2020,Lai2021,Du2021,Duan2022,Itahashi2022,Mook2018,Kondo2022}. However, its counterpart induced by the real-space topology has not been reported till now. It is well known that the geometric phase derived from the adiabatic evolution is crucial for the Hall transport. Notably, in the three-wave mixing process, the accumulation of adiabatic geometric phase takes place not only on incident waves but also on nonlinear ones \cite{Tymchenko2015,Li2015,Li2020}. One thus expects that magnons generated in the nonlinear three-magnon process in spin textures \cite{Wang2021,Wang2022,Schultheiss2009,Iwasaki2014,Zhang2018,Schultheiss2019,Korber2020} may also experience a topological Hall effect subject to the conventional gauge field, but it is not clear whether the rest two gauge fields play any role.

\begin{figure}
  \centering
  \includegraphics[width=0.48\textwidth]{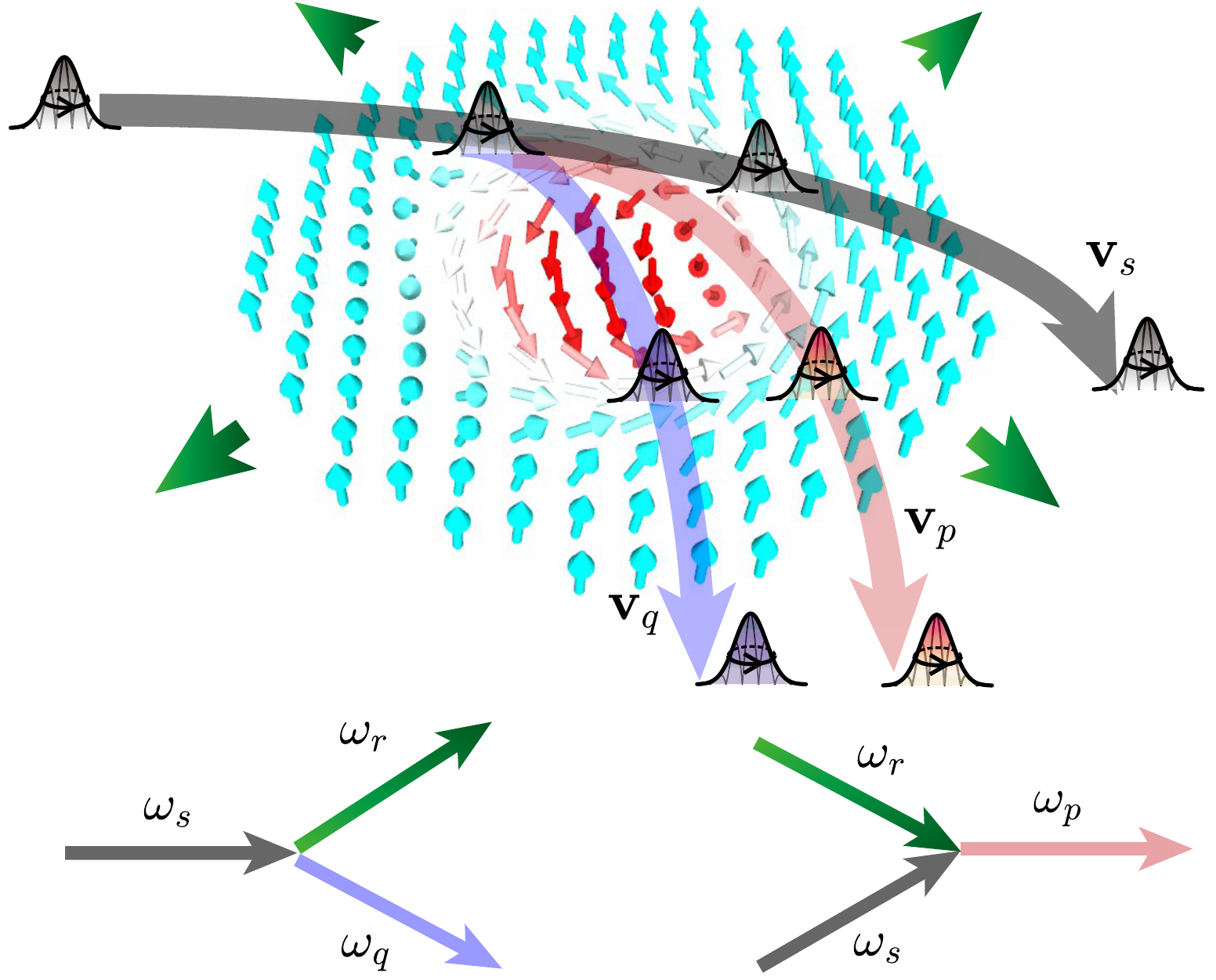}\\
  \caption{Schematic illustration of the nonlinear topological magnon spin Hall effect in magnon-AFM skyrmion scattering. Circles with arrows indicate the handedness of AFM magnons. Incident, skyrmion breathing, sum-frequency, and difference-frequency modes are denoted by black, green, red, and blue colors, respectively. ${\bf v}_{s,p,q}$ represent the velocity of three propagating magnon wavepackets. The bottom panel shows the magnon splitting (left) and confluence (right) processes. It is noted that magnons with opposite handedness will experience magnitude-equal but opposite Lorentz forces, resulting in the opposite transverse displacement (not shown).}\label{fig1}
\end{figure}

In this Letter, we aim to reveal the concealed gauge fields by addressing the nonlinear Hall transport of magnons in spin textures. To this end, we theoretical study the nonlinear interaction between polarized magnons and magnetic skyrmions in antiferromagnets. We show that the two long-sought gauge fields are actually hidden in the nonlinear magnon transport. By analyzing the ``bunny ears'' scattering pattern of three-magnon processes between the circularly-polarized magnon and breathing skyrmion in an antiferromagnet, we discover a giant Hall angle of both confluence and splitting modes. The Hall angle reverses its sign when one switches the handedness of incident AFM magnons. We dub it nonlinear topological magnon spin Hall effect. Our findings are deeply connected to both the nonconservation of magnon number and the spin-texture-induced Berry curvature in real space, as shown in Fig. \ref{fig1}.

\begin{figure}
  \centering
  \includegraphics[width=0.48\textwidth]{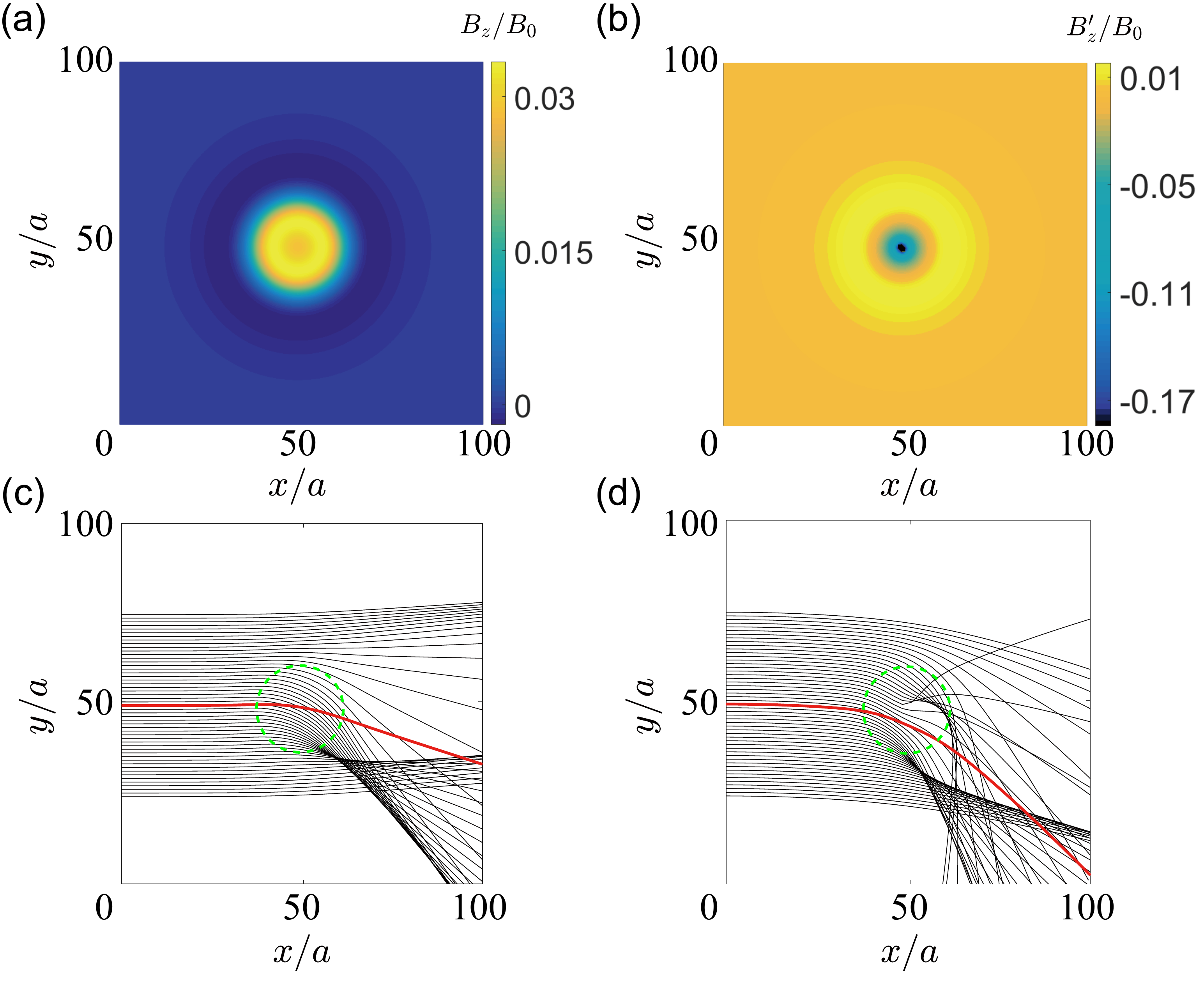}\\
  \caption{Spatial distribution of dimensionless field $B_z/B_0$ (a) and $B'_z/B_0$ (b). Spin wave trajectories in real space under fictitious magnetic field ${\bf B}$ (c) and ${\bf B}+{\bf B}'$ (d), where different black
curves represent trajectories of magnon wavepackets with different impact parameters, the red curve
indicates the averaged trajectory of 51 magnon wavepackets, and the
dashed green circle labels skyrmion's wall center ($l_{z}=0$).}\label{fig2}
\end{figure}

\textit{Model.---}Let us consider a chiral antiferromagnet described by the following Lagrangian \cite{Kim2019}
\begin{equation}\label{Eq1}
\begin{aligned}
\mathcal{L}=\int(\partial_t{\bf l})^2d{\bf r}-\mathcal{H},
\end{aligned}
\end{equation}
where ${\bf l}$ is the normalized N$\rm{\acute{e}}el$ vector and $\mathcal{H}=\int\big [J(\nabla{\bf l})^2+D{\bf l}\cdot(\nabla\times{\bf l})-Kl_z^2\big ]d{\bf r}$ is the system Hamiltonian including the exchange energy, Dzyaloshinskii-Moriya interaction (DMI), and magnetic anisotropy, with $J$, $D$, and $K$ being the exchange stiffness, DMI strength, and anisotropy coefficient, respectively. To facilitate the analysis, we use the $3\times3$ matrix $\mathcal{R}=$exp$(\phi L_z)$exp$(\theta L_y)$ to rotate the ${z}-$axis to the equilibrium direction of the stagger vector ${\bf l}_0$, i.e., $\mathcal{R}{\bf e}_z={\bf l}_0=(\sin\theta\cos\phi,\sin\theta\sin\phi, \cos\theta)$ with the polar angle $\theta$ and azimuthal angle $\phi$. Here, $L_z$ and $L_y$ are generators of the three-dimensional rotations about the $z$ and $y$ axis, respectively.
To investigate the magnon excitation and transport in spin textures, we introduce the magnon creation ($a^{\dagger}$) and annihilation ($a$) operators by the Holstein-Primakoff transformation on the vector ${\bf n}=\mathcal{R}^{-1}{\bf l}$ \cite{HP}. We expand the bosonic operator as $a=a_se^{i{\bf k}_s\cdot{\bf r}}+a_pe^{i{\bf k}_p\cdot{\bf r}}+a_qe^{i{\bf k}_q\cdot{\bf r}}+a_r\psi_r$, where $a_s, a_p, a_q,$ and $ a_r$ are operators of incident magnon, sum-frequency, difference-frequency, and the skyrmion breathing modes \cite{Mochizuki2012}, respectively. ${\bf k}_s,$ ${\bf k}_p,$ and ${\bf k}_q$ are the corresponding wave vectors of three propagating modes in the far-field region, and $\psi_r$ is the wavefunction of the localized breathing mode. Furthermore, we assume that the magnon excitation is in the form of a wave packet, which has a fixed shape and can be described by its central position ${\bf r}_i(t)$, with $i = s, p, q$. In terms of these collective coordinates \cite{Thiele}, the Lagrangian can be simplified as a function of the position ${\bf r}_i$ and the group velocity ${\bf v}_i=\dot{\bf r}_i$ of the magnon wavepacket \cite{Lan2021}. Keeping up to third-order terms, the total Hamiltonian can be expressed as $\mathcal{H}=\mathcal{H}_2+\mathcal{H}_3$. Here, the quadratic Hamiltonian is
\begin{equation}\label{Eq2}
\mathcal{H}_2=\sum_{i=s,p,q}2a_i^{\dagger}a_i\int \Big [{\frac{1}{J}\omega_i^2{\bf v}_i^2}-(2{\bf A}_{12}+\frac{D{\bf l}_0}{J})\cdot\omega_i{\bf v}_i \Big ]d{\bf r},\\
\end{equation}
which determines the magnon dispersion relation. The cubic Hamiltonian $\mathcal{H}_3=\mathcal{H}_{3s}+\mathcal{H}_{3p}+\mathcal{H}_{3q}$ includes contributions from the incident term $\mathcal{H}_{3s}$, sum-frequency term $\mathcal{H}_{3p}$, and difference-frequency term $\mathcal{H}_{3q}$
\begin{equation}\label{Eq3}
\begin{aligned}
\mathcal{H}_{3s}=&\int\omega_s{\bf v}_s\cdot\bigg \{-\frac{1}{\sqrt{2}}\big [(i{\bf A}_{13}+{\bf A}_{23})+\frac{D}{2J}(i{\bf e}_{\phi}+{\bf e}_{\theta})\big ]\\
&\big [3a_qa_ra_s^{\dagger}e^{i(-{\bf k}_s+{\bf k}_q)\cdot{\bf r}}+a_p^{\dagger}a_ra_se^{i({\bf k}_s-{\bf k}_p)\cdot{\bf r}}\big ]\psi_r+{\rm H.c.}\bigg\}d{\bf r},\\
\mathcal{H}_{3p}=&\int\omega_p{\bf v}_p\cdot\bigg\{-\frac{3}{\sqrt{2}}\big [(i{\bf A}_{13}+{\bf A}_{23})+\frac{D}{2J}(i{\bf e}_{\phi}+{\bf e}_{\theta})\big ]\\
&a_sa_ra_p^{\dagger}e^{i({\bf k}_s-{\bf k}_p)\cdot{\bf r}}\psi_r+{\rm H.c.}\bigg\}d{\bf r},\\
\mathcal{H}_{3q}=&\int\omega_q{\bf v}_q\cdot\bigg\{-\frac{1}{\sqrt{2}}\big [(i{\bf A}_{13}+{\bf A}_{23})+\frac{D}{2J}(i{\bf e}_{\phi}+{\bf e}_{\theta})\big ]\\
&a_sa_r^{\dagger}a_q^{\dagger}e^{i({\bf k}_s-{\bf k}_q)\cdot{\bf r}}\psi_r^*+{\rm H.c.}\bigg\}d{\bf r},\\
\end{aligned}
\end{equation}
where ${\bf A}_{\nu\nu'}=\mathcal{A}_{x,\nu\nu'}{\bf e}_x+\mathcal{A}_{y,\nu\nu'}{\bf e}_y$ are the gauge fields ($\nu,\nu'=1,2,3$), ${\bf e}_{\theta}$ and ${\bf e}_{\phi}$ are two unit vectors in spherical coordinates, and $\omega_s,$ $\omega_p,$ and $\omega_q$ are, respectively, the frequencies of the incident, confluence, and splitting magnons meeting law of energy conservation, i.e., $\omega_{p(q)}=\omega_{s}\pm\omega_r$ with $\omega_r$ the skyrmion breathing frequency, see the bottom panel of Fig. \ref{fig1}. Equations \eqref{Eq2} and \eqref{Eq3} show that the conventional gauge field ${\bf A}_{12}$ only appears in the quadratic term, while gauge fields ${\bf A}_{13}$ and ${\bf A}_{23}$ emerge in the nonlinear three-magnon processes. To reveal their role in the magnon transport, we employ the Euler-Lagrangian formula to derive equations of motion of magnon wavepackets
\begin{equation}\label{Eq4}
\begin{aligned}
&a_i^{\dagger}a_i\frac {\hbar\omega_i^2}{eJ}\dot{{\bf v}_i}-a_i^{\dagger}a_i\omega_i{\bf v}_i\times {\bf B}-{\bf F}_i^{{\rm cubic}}=0,(i=s,p,q),\\
\end{aligned}
\end{equation}
where ${\bf B}=B_z{\bf e}_z$ with $B_z=\frac{\hbar}{e}\big [{\bf \nabla} \times ({\bf A}_{12}+\frac{D{\bf l}_0}{2J})\big ]_z=\frac{\hbar}{e}\big [{\bf l}_0\cdot(\partial_x{\bf l}_0\times\partial_y{\bf l}_0)+({\bf \nabla} \times\frac{D{\bf l}_0}{2J})_z\big ]$ is the conventional fictitious magnetic field for the linear magnon transport with the reduced Planck constant $\hbar$ and elementary charge $e$. See Supplemental Material \cite{SM} for the derivation of \eqref{Eq4}. It is noted that we have ignored the effective electric field associated with the skyrmion static energy due to its negligible role in the magnon Hall effect. The extra force ${\bf F}_i^{{\rm cubic}}$ originates from the three-magnon process and the newfound gauge fields, with the following expression
\begin{equation}\label{Eq5}
\begin{aligned}
&{\bf F}_i^{{\rm cubic}}=c_i{\bf v}_i\times {\bf B}',(i=s,p,q),\\
\end{aligned}
\end{equation}
where ${\bf B}'=B_z'{\bf e}_z$ with $B_z'=\frac{\hbar}{e}({\bf \nabla} \times {\bf A}_{23})_z+\frac{\hbar D}{2Je}({\bf \nabla} \times {\bf e}_{\theta})_z=\frac{\hbar}{e}\big [\partial_y{\bf l}_0\cdot\partial_x({\bf n}\times\frac{{\bf e}_z}{\sin\theta})-\partial_x{\bf l}_0\cdot\partial_y({\bf n}\times\frac{{\bf e}_z}{\sin\theta})\big ]+\frac{\hbar D}{2Je}(\nabla\times\frac{{\bf e}_z+\cos\theta{\bf l}_0}{\sin\theta})_z$ represents the new fictitious magnetic field playing a role merely when the nonlinear three-magnon process occurs. Due to the circular symmetry of skyrmion, the ${\bf A}_{13}$ component is absent. Here, $c_s=\frac{\omega_s}{4}(g_pa_p^{\dagger}a_ra_s+3g_qa_q^{\dagger}a_r^{\dagger}a_s+{\rm H.c.})$, $c_p=\frac{3\omega_p}{4}(g_pa_sa_ra_p^{\dagger}+{\rm H.c.})$, and $c_q=\frac{\omega_q}{4}(g_pa_sa_r^{\dagger}a_q^{\dagger}+{\rm H.c.})$, with overlap integrals $g_p=\frac{1}{\sqrt{2}V}\int e^{i({\bf k}_s-{\bf k}_p)\cdot{\bf r}}\psi_rd{\bf r}$, $g_q=\frac{1}{\sqrt{2}V}\int e^{i({\bf k}_s-{\bf k}_q)\cdot{\bf r}}\psi_r^*d{\bf r}$, and $V$ being the system volume. As shown in Eq. \eqref{Eq4}, the spin-wave packet can be regarded as a particle-like object moving in its own parameter space \cite{Sundaram1999} subject to fictitious magnetic fields (${\bf B}$ and ${\bf B}'$). The first term on the left-hand side of Eq. \eqref{Eq4} characterizes the acceleration of magnons. The second term represents the effective Lorentz force from the quadratic Hamilton $\mathcal{H}_2$, resulting in the conventional topological magnon Hall effect. Interestingly enough, the third term induces an extra Lorentz force on the wavepacket, leading to the nonlinear topological magnon Hall effect. The spatial distributions of the dimensionless magnetic fields $B_z/B_0$ and $B'_z/B_0$ are shown in Figs. \ref{fig2}(a) and \ref{fig2}(b), respectively, where $B_0=\hbar/{a^2e}$ with $a$ being the lattice constant. It is noted that $B_0\approx 660$ T for $a=1$ nm. Due to the rotational symmetry of the Bloch skyrmion, both magnetic fields ${\bf B}/B_0$ and ${\bf B}'/B_0$ have the circular symmetry. For ${\bf B}/B_0$, its main origin comes from the topological charge density of skyrmion, and the total magnetic flux is 4$\pi$ \cite{Volovik1987}. The spatial distribution of ${\bf B}'/B_0$, however, is similar to the fictitious magnetic field distribution of the target skyrmion \cite{Tang2021,Ndiaye2017} with a vanishing total flux but a singularity at the skyrmion core.

\textit{Revealing the concealed fictitious magnetic field.---}In nonlinear magnon-skyrmion scatterings, the time-evolution of populations of confluence and splitting modes is governed by the coupled Heisenberg equations: $i\dot{a}_p=(\Delta_p-i\alpha\omega_p)a_p+ \tilde{g}_pa_sa_r$ and $i\dot{a}_q=(\Delta_q-i\alpha\omega_q)a_q+\tilde{g}_qa_sa_r^{\dagger}$. Here, the detuning parameter $\Delta_{p(q)}=\omega_{p(q)}-\omega_0$ with the driving microwave frequency $\omega_0$, $\tilde{g}_p=\int\big [-2g_{1,\mu}ik_{p,\mu}\psi_r+g_{2,\mu}^*(\partial_{\mu}\psi_r+ik_{s,\mu})-\frac{5}{\sqrt2}K\sin \theta\cos\theta\big ]e^{i(\bf{k}_s-\bf{k}_p)\cdot\bf{r}}d\bf{r}$ and $\tilde{g}_q=\int\big [g_{2,\mu}(\partial_{\mu}\psi_r^*+ik_{q,\mu})+2g_{1,\mu}^*ik_{s,\mu}\psi_r^*-\frac{5}{\sqrt2}K\sin \theta\cos\theta\big ]e^{i(\bf{k}_s-\bf{k}_q)\cdot\bf{r}}d\bf{r}$ where the Einstein summation rule is applied, coefficients $g_{1,\mu}=\frac{3J}{2\sqrt{2}}(\mathcal{A}_{\mu,13}-i\mathcal{A}_{\mu,23})+\frac{3D}{4\sqrt{2}}(e_{\phi,\mu}-ie_{\theta,\mu})$ and $g_{2,\mu}=\frac{J}{\sqrt{2}}(-\mathcal{A}_{\mu,13}-i\mathcal{A}_{\mu,23})+\frac{D}{2\sqrt{2}}(-e_{\phi,\mu}-ie_{\theta,\mu})$ denote the strength of the three-magnon confluence and splitting, respectively, and $\alpha$ is the Gilbert damping constant. Then, one can analytically derive the steady-state magnon populations as $a_p=\frac{ga_sa_r}{\epsilon+i\alpha(\omega_s+\omega_r)}$ and $a_q=\frac{ga_sa_r^{\dagger}}{\epsilon-i\alpha(\omega_s-\omega_r)}$ with $\epsilon=\omega_s-\omega_r$. Here, we have adopted the approximation $\tilde{g}_p\approx \tilde{g}_q\approx g$ which is justified by the small difference between confluence and splitting frequencies since $\omega_r\ll\omega_s$. We therefore obtain
\begin{equation}\label{Eq6}
\begin{aligned}
&m_{{\rm sw},\,i}\dot {\bf v}_i-e{\bf v}_i\times \sigma({\bf B}+\lambda_i{\bf B}')=0, (i=s,p,q),\\
\end{aligned}
\end{equation}which is the main result of this work (see Supplemental Material \cite{SM} for detailed derivations). Here, $m_{{\rm sw},i}=\hbar\omega_i/J$ is the effective mass of the spin-wave packet in antiferromagnets, $\sigma=\mp1$ represents the left/right-hand magnon polarizations, $\lambda_s=n_r(\frac{gg_p}{4\epsilon}+\frac{3gg_q}{4\epsilon}+{\rm H.c.})$, $\lambda_p=\frac{3}{4}(\frac{\epsilon g_p}{g}+{\rm H.c.})$, and $\lambda_q=\frac{1}{4}(\frac{\epsilon g_q}{g}+{\rm H.c.})$ with the particle number of skyrmion breathing mode $n_r=\langle a_r^{\dagger}a_r\rangle$. Equation \eqref{Eq6} shows that the extra effective Lorentz force $e\lambda_i{\bf v}_i\times \sigma{\bf B}'$ $(i=s,p,q)$ is mode-dependent. For incident magnons, the extra Lorentz force is proportional to the product of the skyrmion breathing number $n_r\ (\ll1)$, the coupling parameter $g/\epsilon$, and the overlap integral $g_{p,q}$. In general, magnon populations of confluence and splitting modes are far less than the incident one. It implies $g/\epsilon,\ g_{p,q}\ll1$. The effect of ${\bf B}'$ on incident magnons is thus negligible. However, for the confluence and splitting modes, parameters $\lambda_{p,q}$ are inversely proportional to $g/\epsilon$, the additional effective Lorentz force is therefore expected to bring an enormous effect.

To explore the role of fictitious magnetic fields on the magnon transport, we numerically solve Eq. \eqref{Eq6} both without and with the new fictitious magnetic field ${\bf B}'$. In calculations, we consider a right-handed magnon wavepacket ($\sigma=1$), and set the incident-magnon's initial velocity $v_{s}(t=0) = 2.65$ (with unit $J/a\omega$) along $x$ direction, magnon mass $m_{\rm sw}=0.31$ (with unit $\hbar\omega/J$), and coefficient $\lambda_{s,p,q}=1$. It is observed that the effective magnetic field ${\bf B}'$ significantly enhances the magnon Hall effect, as displayed in Figs. \ref{fig2}(c) and \ref{fig2}(d). Due to the singularity of the fictitious magnetic field ${\bf B}'$, we note anomalous magnon trajectories near the skyrmion center. Below, we verify our theoretical predictions by full micromagnetic simulations using MUMAX3 package \cite{Vansteenkiste2014}.

We consider an AFM thin film of dimension $1000$ $\times1000$ $\times1$ nm$^3$, hosting a Bloch-type skyrmion at the film center \cite{Binz2009,Yu2010,Huang2013}. Magnetic parameters of ${\rm KMnF_3}$ \cite{Barker2016}: $J=6.59$ pJ/m, $K=1.16\times10^{5}$ J/m$^3$, and $D=1$ mJ/m$^2$ are used in the simulations, which gives rise to a skyrmion radius $\approx11$ nm (defineds as the radius of circle $l_z=0$). The Gilbert damping is set as $\alpha=0.001$. To efficiently generate polarized magnons and the three-wave mixing, we apply a microwave field ${\bf h}_{\rm RH/LH}(t) = h_0 [\cos(\omega_st){\bf e}_x\mp\sin(\omega_st){\bf e}_y]$ with amplitude $h_0 = 50$ mT and frequency $\omega_s/2\pi = 1.205$ THz (generating the incident magnon of wavelength $\approx 15.2$ nm) on one sublattice in a narrow region: $-401$ nm$\leq x\leq -399$ nm and a local field ${\bf h}_r(t) = h_r \sin(\omega_rt){\bf e}_z$  over the skyrmion with amplitude $h_r = 5$ mT and $\omega_r/2\pi = 0.095$ THz (the skyrmion breathing frequency) \cite{SM}. Here, RH and LH represent the abbreviation of microwave with right and left handedness, respectively. Absorbing boundary conditions are adopted to eliminate the spin-wave reflection by film edges \cite{Venkat2018}.
\begin{figure}
  \centering
  \includegraphics[width=0.48\textwidth]{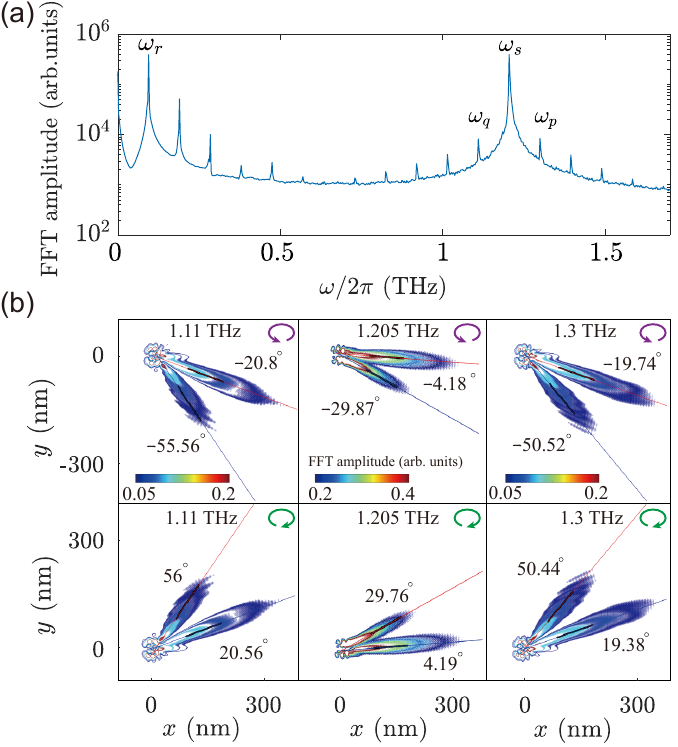}\\
  \caption{(a) MFC in the nonlinear scattering between the incident magnon and AFM skyrmion. (b) Isoline maps for righted-handed (top panel) and left-handed (bottom panel) magnons scattered by the skrymion at the origin. In each panel, modes from left to right correspond to splitting, incident, and confluence magnons, respectively.}\label{fig3}
\end{figure}

To analyze the magnon spectrum in the skyrmion area, we implement fast Fourier transform of local magnetic moments. Figure \ref{fig3}(a) shows the emerging magnon frequency comb (MFC) \cite{Hula2022} in the terahertz region, where the mode spacing of the comb is exactly the skyrmion breathing frequency. Furthermore, we plot the isoline map of incident, confluence, and splitting modes to analyze the Hall angle of each mode, as shown in Fig. \ref{fig3}(b). We observe an interesting ``bunny ears'' pattern of magnons scattering off the AFM skyrmion, with red and blue lines denoting the propagation direction of two branches. Here, the Hall angle is defined as the included angle between each branch and the horizontal line. Compared with the incident mode, the Hall angle of nonlinear modes nearly doubles (quintuples) for the main (secondary) branch of the ``bunny ears'' [see Fig. \ref{fig3}(b)], where the major (secondary) branch is referred to as the one with a large (small) Hall angle. More importantly, by flipping the chirality of incident magnons, we observe an opposite magnon Hall motion [comparing the top and bottom panels in Fig. \ref{fig3}(b)]. The small difference of Hall angles between right-handed and left-handed magnons results from the dipolar field \cite{SM}. Micromagnetic simulations thus offer solid evidences for the nonlinear topological magnon spin Hall effect as we predicted above.

\begin{figure}
  \centering
  \includegraphics[width=0.48\textwidth]{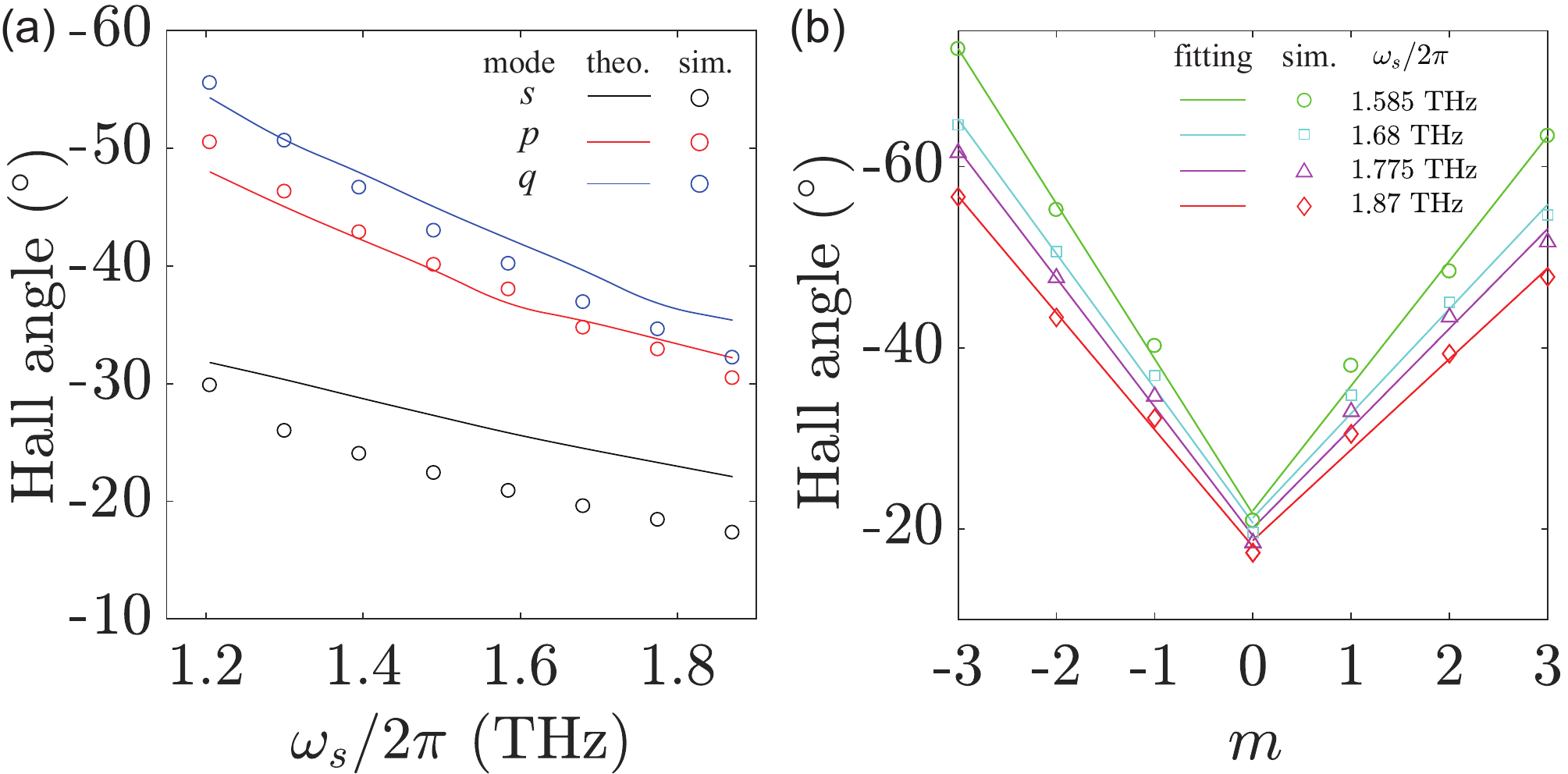}\\
  \caption{(a) The Hall angle of the main branch of ``bunny ears'' as a function of the driving frequency for the incident (black), confluence (blue), and splitting (red) modes. Symbols are micromagnetic simulations and curves are analytical fitting by solving Eq. \eqref{Eq6}. (b) The Hall angle as a function of the mode index $m$ for different incident magnon frequencies. Symbols and lines represent micromagnetic simulations and linear fittings, respectively.}\label{fig4}
\end{figure}

Furthermore, we derive the frequency-dependent Hall angle by fitting the flow direction of the main branch of the isosurface of each mode. Figure \ref{fig4}(a) plots the quantitative comparison between theoretical calculations and micromagnetic simulations for incident (black), confluence (blue), and splitting (red) modes. It shows that the Hall angle monotonically decreases with the increase of the mode frequency. Simulation results can be well explained by the analytical model \eqref{Eq6}  with parameters $n_r=0$, $g=49$ MHz, and $g_p=\frac{1}{3}g_q=9.4\times10^{-6}$. A vanishing mode number of skyrmion breathing is justified by its small wave amplitude. The coupling coefficient $g$ is independently obtained by numerically solving the coupled Heisenberg equations. Acceptable deviations could be attributed to the simplified wavepacket treatment in the present formalism and the neglected topological electric-field component of the gauge fields. Figure \ref{fig4}(b) shows the Hall angle of nonlinear magnons over a broad frequencies $\omega_s+m\omega_r$ in the MFC with integer $m$ labeling the order of the spectrum line. Their ``bunny ears'' scattering patterns are plotted in Supplemental Material \cite{SM}. It is found that the noninear Hall angle increases linearly with $|m|$. This monotonic dependence is reminiscent of the refraction process of light waves through multilayer media \cite{Multiple}, where the refraction angle accumulates upon each scattering layer. It is noted that the slope of the linear trendline decreases as the incident magnon's frequency $\omega_s$ increases.

\textit{Discussion.---}In the above calculations, we have considered rotationally symmetric spin textures, the ${\bf A}_{13}$ gauge field thus vanishes. However, the curl of gauge field induced by DMI becomes finite when the rotational symmetry is broken. We then envision contributions from ${\bf A}_{13}$ in generating the fictitious magnetic field for elliptical skyrmions \cite{Jena2020}.

To summarize, we revealed the long-sought gauge fields concealed in the nonlinear magnon transport. By investigating the three-wave mixing between propagating magnons and breathing skyrmions, we found giant Hall angles emerging for each nonlinear spectrum line of the MFC. We further identified that the sign of the Hall angle is reversed by switching the chirality of incident magnons, and we dub it nonlinear topological magnon spin Hall effect. Our findings are intimately connected to the particle number nonconservation of magnons and thus applicable to generic bosons, which does not have the low-energy fermionic counterpart. Our results significantly advance the understanding of the nonlinear Hall effect and pave the way to probing the gauge field by frequency comb.

\begin{acknowledgments}
This work was funded by the National Key Research Development Program under Contract No. 2022YFA1402802 and the National Natural Science Foundation of China (NSFC) (Grant No. 12074057). Z.W. acknowledges financial support from the NSFC (Grant No. 12204089) and the China Postdoctoral Science Foundation under Grant No. 2019M653063. H.Y.Y. acknowledges the European Union's Horizon 2020 research and innovation programme under Marie Sk{\l}odowska-Curie Grant Agreement SPINCAT No. 101018193.
\end{acknowledgments}

\end{document}